\documentclass[floatfix,twocolumn,showpacs,preprintnumbers,amsmath,amssymb,prl,superscriptaddress]{revtex4-1}

\newcommand{\N}{$\hat{\rho_1}$ }
\newcommand{\Ne}{$\hat{\rho_1}$}
\newcommand{\CF}{$n_0$ }
\newcommand{\CFe}{$n_0$}
\usepackage{graphicx}
\usepackage{epstopdf}
\usepackage{dcolumn}
\usepackage{color}
\usepackage{multirow}
\usepackage{float}

\graphicspath{{./imgs/}{./}}

\begin{document}
\preprint{Draft version \today}

\title{Revealing the Condensate and Non-Condensate Distributions in the Inhomogeneous Bose-Hubbard Model}

\author{Ushnish Ray and David M. Ceperley}
\affiliation{Department of Physics, University of Illinois at Urbana-Champaign, Urbana, IL 61801, USA}
	
\begin{abstract}
We calculate the condensate fraction and the condensate and non-condensate spatial and momentum distribution of the Bose-Hubbard model in a trap. From our results, it is evident that using approximate distributions can lead to erroneous experimental estimates of the condensate. Strong interactions cause the condensate to develop pedestal-like structures around the central peak that can be mistaken as non-condensate atoms. Near the transition temperature, the peak itself can include a significant non-condensate component. Using distributions generated from QMC simulations, experiments can map their measurements for higher accuracy in identifying phase transitions and temperature. 
\end{abstract}


\maketitle
The Bose-Hubbard (BH) model has been the focus of intensive research over the past decade as a prototypical example of strongly correlated physics, especially since the model was realized with cold atoms in optical lattices \cite{Greiner2002}. Optical lattice experiments (OLE) are prime candidates for quantum simulations due to their extensive tunability and ease of control. Therefore they serve as ideal systems to study dynamical phenomena of many body effects in strongly interacting systems. Before studies can be meaningful, however, systematic characterization of equilibrium properties are crucial, the most important quantities being temperature (T) and density. However, direct or in-situ measurements of temperature in OLEs is an area of active research \cite{mckay09,mckay_thesis}. Further, the study of phase transitions  requires a good understanding of an observable that can be used as a probe for the state of the system. Theoretically, the natural choice is the order parameter, which for the BH model is the condensate fraction $n_0=N_0/N$ where $N_0$ is the total number of condensed atoms and $N$ the  number of atoms. Alternatively, the superfluid fraction \cite{ceperley95, leggett06} characterizes the transition, but this is not simple to measure in cold atom systems. The most easily accessible observables in experiments are the entropy and \CF that come from time-of-flight (TOF) measurements. The former is measured from TOF measurements while atoms are in the harmonic trap (without the lattice) and are then isentropically transferred into the lattice \cite{gericke07, pasienski10, mckay_thesis}. The latter could be measured directly from TOF expansions after all fields are "snapped off." The \CF is particularly useful since, combined with entropy, it could be used for thermometry in experiments \cite{mckay09,trotzky10}.      

In homogeneous systems, \CF is given by a delta function at the origin in momentum space. In TOF images, therefore, its signal would be visible from the appearance of a sharp peak. The presence of the parabolic trapping potential in OLEs, however, renders the system inhomogeneous and \CF is no longer simply given by the occupation number at zero momentum.

The most common modelling approach to handle the trap is mean field (MF) theory: e.g. the Hartree-Fock-Bogoliubov-Popov approximation \cite{rey03,Yi07, anderson04} for small interaction strengths (U) or the site-decoupled approach for large U \cite{oosten01}, together with the local density approximation (LDA). In situations where MF fails or quantitative comparison to measurement is important, we can resort to exact quantum Monte Carlo (QMC) techniques. QMC has been used to directly compare observables with measurements. However, to the best of our knowledge, the order parameter has not been computed and compared directly to experiment \cite{footnote1}.

In the most common approach, experimental TOF images are heuristically fit to obtain the number of condensed atoms under the peaks and the remaining non-condensed atoms. The ratio of the former to the sum of two is defined as the peak weight \cite{trotzky10} or the coherence fraction \cite{mckay_thesis} ($f_0$) that serves as a proxy for \CFe. 
In previous experiments, thermometry was done by comparing full momentum distributions together with $f_0$, peak width ($w_0$) and the visibility directly to QMC results \cite{trotzky10}. The last three observables were further used to characterize the critical temperature ($T_c$) for the transition from the normal to the superfluid phase. Unfortunately, these probes are not necessarily reliable estimates of the order parameter since the relation between \CF and $f_0$ is not well understood. Previous comparisons show large differences \cite{mckay_thesis}. This is unfortunate because \CF is a simple probe for this transition and is also indicative of the effect of interactions, i.e. the quantum depletion (QD). Combined with the entropy measurements, \CF would be an excellent probe for the temperature. 

The difficulties in characterizing the mapping stems from the fact that the underlying condensate and non-condensate distributions are not well understood in trapped systems. If they were known, then \CF could be estimated by counting the number of atoms in the condensate peak and in the background. (We note that an analytical MFT method has been developed for a homogeneous system \cite{wang09}. Although qualitatively useful, this method finds limited use in real trapped systems where it must rely on LDA.) Another serious problem is the role of interaction during TOF expansion; some have argued that the effect is small since by the time the wavefunctions from two adjacent lattice sites start overlapping, the densities would have dropped dramatically \cite{Prokofev08}. However, others argue that the effects are significant and lead to hydrodynamic effects in TOF images \cite{mckay_thesis}. The crucial quantity that dictates the significance of these effects is the initial density. From two different sets of experimental data, it seems that for low densities (central filling of 1 or less) \cite{trotzky10} interaction effects are small, while at large densities \cite{mckay_thesis} (central filling of 3) they could be significant. Even in the case of non-interacting (NI) expansion, however, the interference pattern - reminiscent of Fresnel diffraction - has to be explicitly accounted for \cite{Prokofev08}.


In this paper, we calculate  \CF and  the spatial and momentum distributions for several low density systems modelled with the BH Hamiltonian:
 \begin{equation}
 H=-t\sum_{\langle ij\rangle}\hat{a}_i^{\dagger}\hat{a}_j+\frac{U}{2}\sum_i\hat{n}_i(\hat{n}_i-1)-\sum_i\tilde{\mu}_i\hat{n}_i,
\label{hamiltonian}
 \end{equation}
where $t$ is the hopping integral between nearest neighbor sites $i$ and $j$, $\hat{a}_i^{\dagger}$ ($\hat{a}_i$) is the Boson creation (annihilation) operator, $\hat{n}_i=\hat{a}_i^{\dagger}\hat{a}_i$ is the number operator, and $U$ is the on-site repulsive interaction. Here, $\tilde{\mu}_i=\mu-V(r_i/a)^2$ includes both the chemical potential term and spherical harmonic confining potential with $a$, the lattice spacing and $V = \frac{1}{2}m\omega^2$ the curvature that is given by the mass ($m$) and trap frequency ($\omega$). Energies are given in atomic recoil energy units: $E_r \sim 167$nK for $^{87}Rb$ and the laser wavelength $\lambda = 800$ nm that is used to create the lattice. For our simulations, lattices are between $70^3$ to $100^3$ with open boundary conditions and the number of particles, $N \sim 58000$ to $64000$.

We calculate the single-particle density matrix $\hat{\rho_1}(i,j) = \langle\hat{a}^{\dagger}_i\hat{a}_j\rangle$ \cite{onsager, ceperley95, leggett06, lewart88} using the stochastic series expansion and the directed loop update algorithm \cite{sandvik99, troyer01, sandvik02}. Then the occupation of the single particle states is defined by $\hat{\rho}_1|\psi_i\rangle = N_i|\psi_i\rangle$, where the largest eigenvalue $N_0$ of \N gives the number of condensed atoms and $|\psi_0\rangle$ is the  condensate wavefunction.  The other atoms $N_{nc} = \sum_{i\neq0} N_i=N - N_0$ are non-condensed atoms.

For large systems, obtaining all the occupation modes is challenging because of the Monte Carlo noise in \N and the complexity of a complete diagonalization of \Ne. However, since the condensate is not fragmented and occupies only one mode in these systems, we use an iterative diagonalization procedure to obtain the spatial condensate wave function, $\psi_0(r)$ and the condensate momentum wave function, $\phi_0(k)=\mathcal{F}[\psi_0(r)]$ where $\mathcal{F}$ is the Fourier transform. For an extensive range of systems that we tested, the method is robust and is able to withstand statistical noise provided we do not enforce Hermitian symmetry; doing so biases the results. We use other symmetries to reduce the noise in \Ne. For a full discussion, we refer to \cite{ray12a}. The spatial non-condensate distribution is then given as $n_{nc}(r) = n(r) - N_0|\psi_0(r)|^2$, where $n(r)$ is the spatial density. The total momentum distribution is
\begin{equation}
n(k)=|w(k)|^2\sum_{jl}e^{ik\cdot(j-l)} \rho(i,j) = n_0(k) + \sum_{p=1} n_p(k).
\label{cmomfraction}
\end{equation}
We have explicitly included the wannier envelope $|w(k)|^2$ that is needed to go back to the continuum model from the lattice BH model. The last term on the rhs is the momentum non-condensate distribution $n_{nc}(k)$. In order to match with experiments, we must include the finite TOF effects \cite{Prokofev08}, by adding an additional site dependent phase term to (\ref{cmomfraction}) so that $n^{\tau}_p(k) = |w(k)|^2N_p\sum_{jl}e^{ik\cdot(j-l) - i(m/2\hbar \tau)(j^2-l^2)}\langle j|\psi_p\rangle\langle\psi_p|l\rangle$, where $\tau$ is the TOF time, $m$ is the particle mass. We use $n^{\tau}_0(k)$ to denote the finite TOF condensate and $n^{\tau}_{nc}(k)$ for the finite TOF non-condensate distributions. 

\begin{figure*}[t]
\begin{tabular}{cccc}
	\multicolumn{2}{c}{$U/t = 25$ $N \sim 60,000$ $\omega = 68.1 Hz$} &
	\multicolumn{2}{c}{$U/t = 40$ $N \sim 58,000$ $\omega = 67.6 Hz$} \\

	\fbox{$k_bT/t=2.456$ $n_0=0.218(2)$} & 
	\fbox{$k_bT/t=0.98$ $n_0=0.626(6)$} & 
	\fbox{$k_bT/t=1.96$ $n_0=0.198(2)$} & 
	\fbox{$k_bT/t=0.98$ $n_0=0.484(6)$} \\	

    \includegraphics[clip,width=39mm, height=25mm, angle=0]{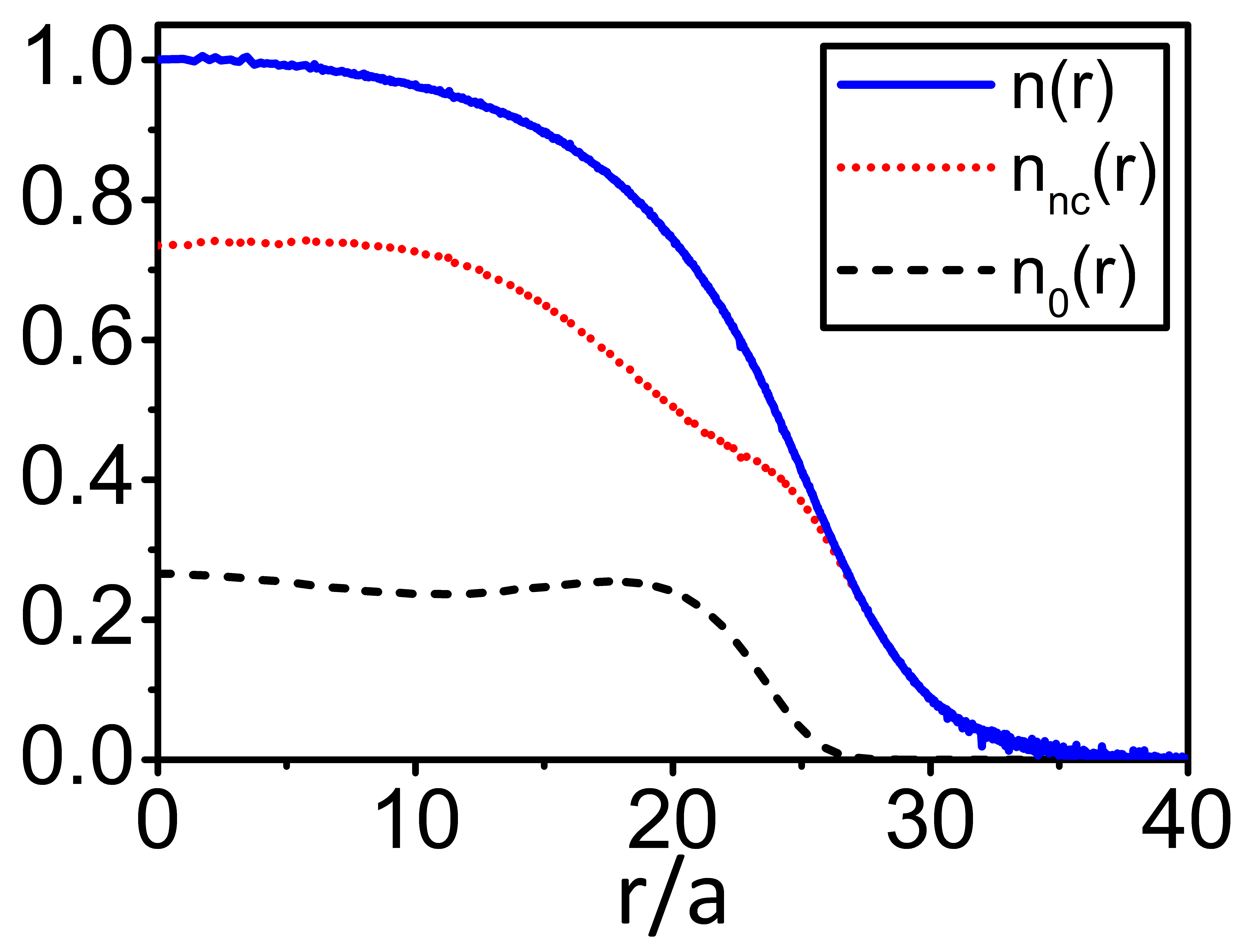} &
    \includegraphics[clip,width=39mm, height=25mm, angle=0]{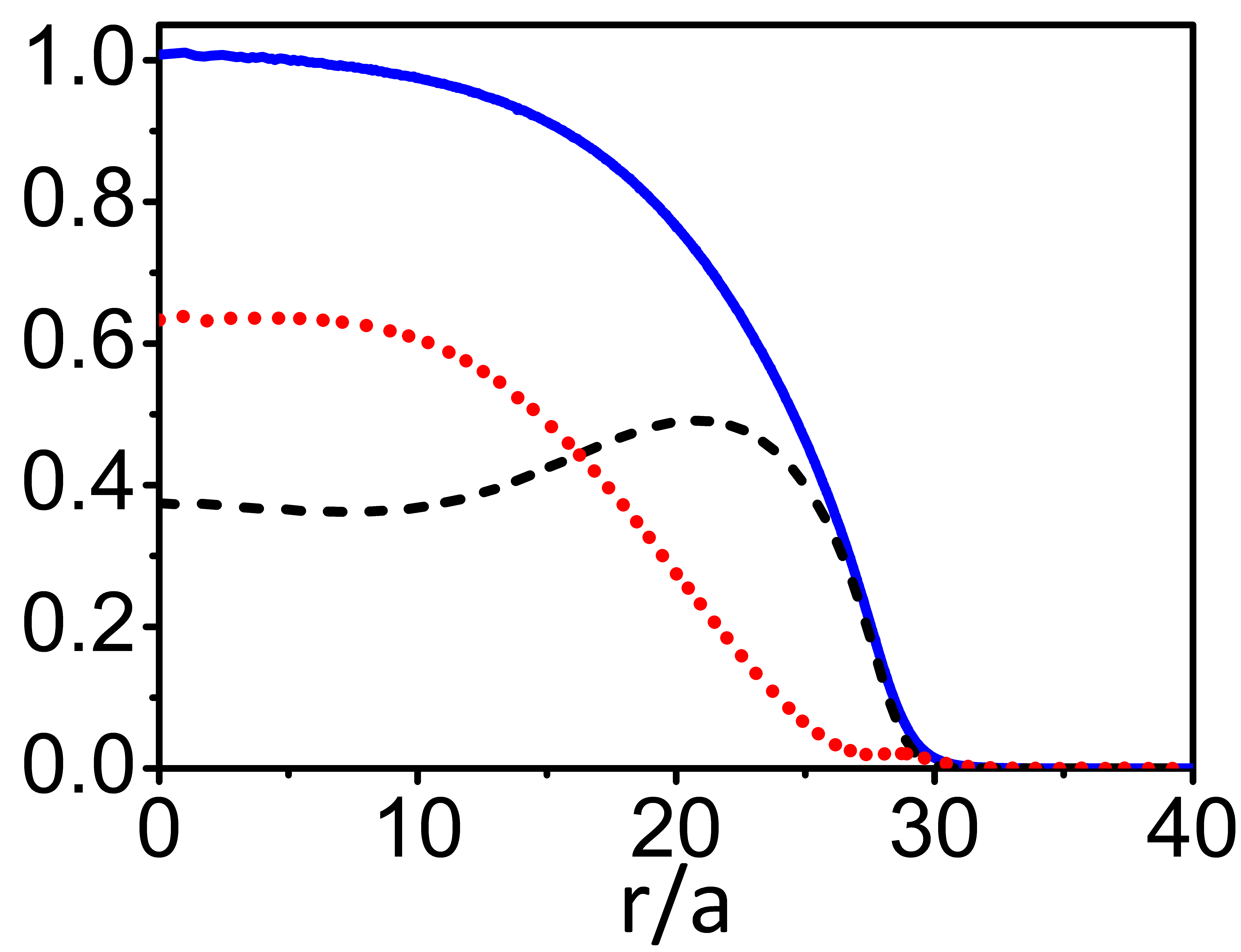} &
    \includegraphics[clip,width=39mm, height=25mm, angle=0]{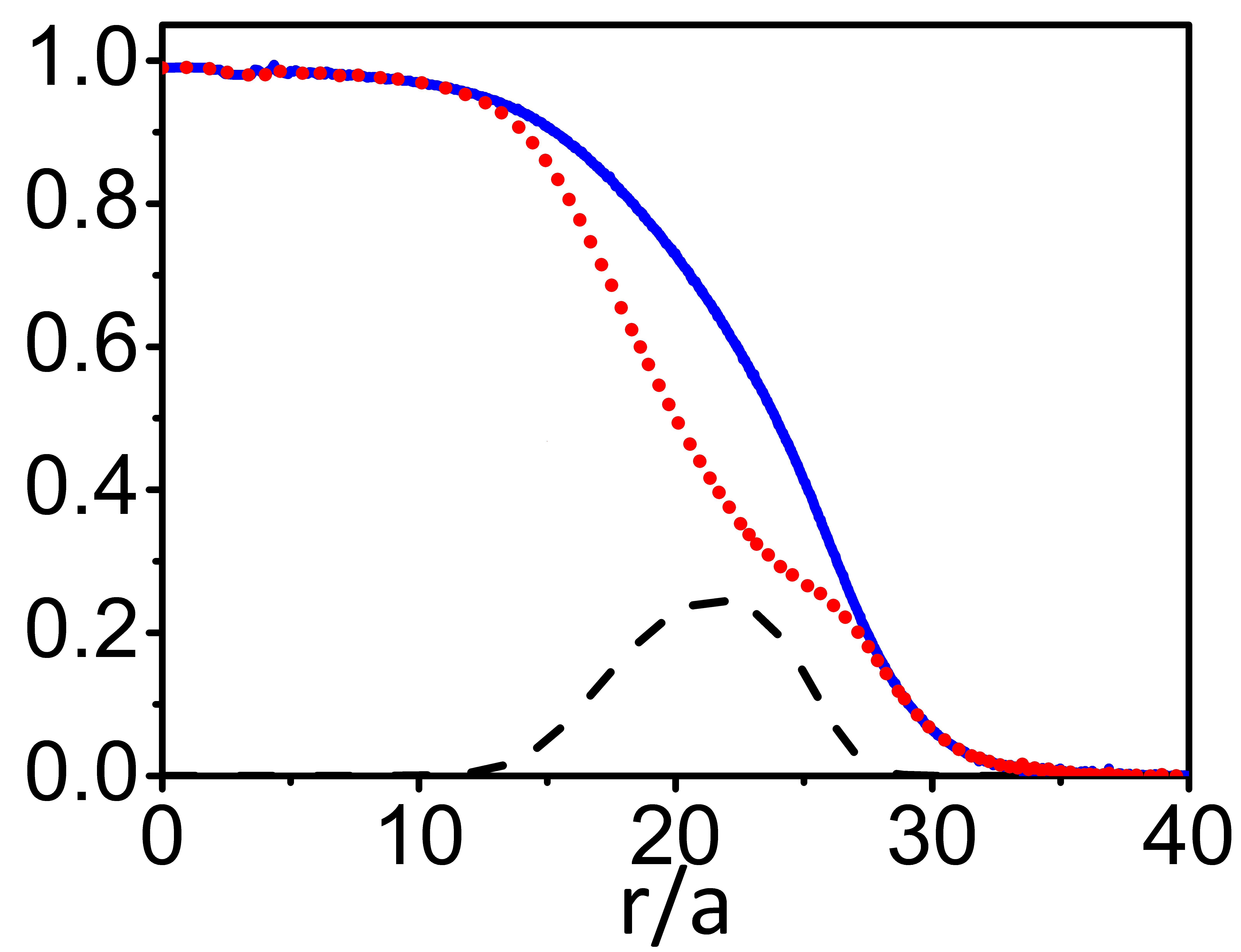} &
    \includegraphics[clip,width=39mm, height=25mm, angle=0]{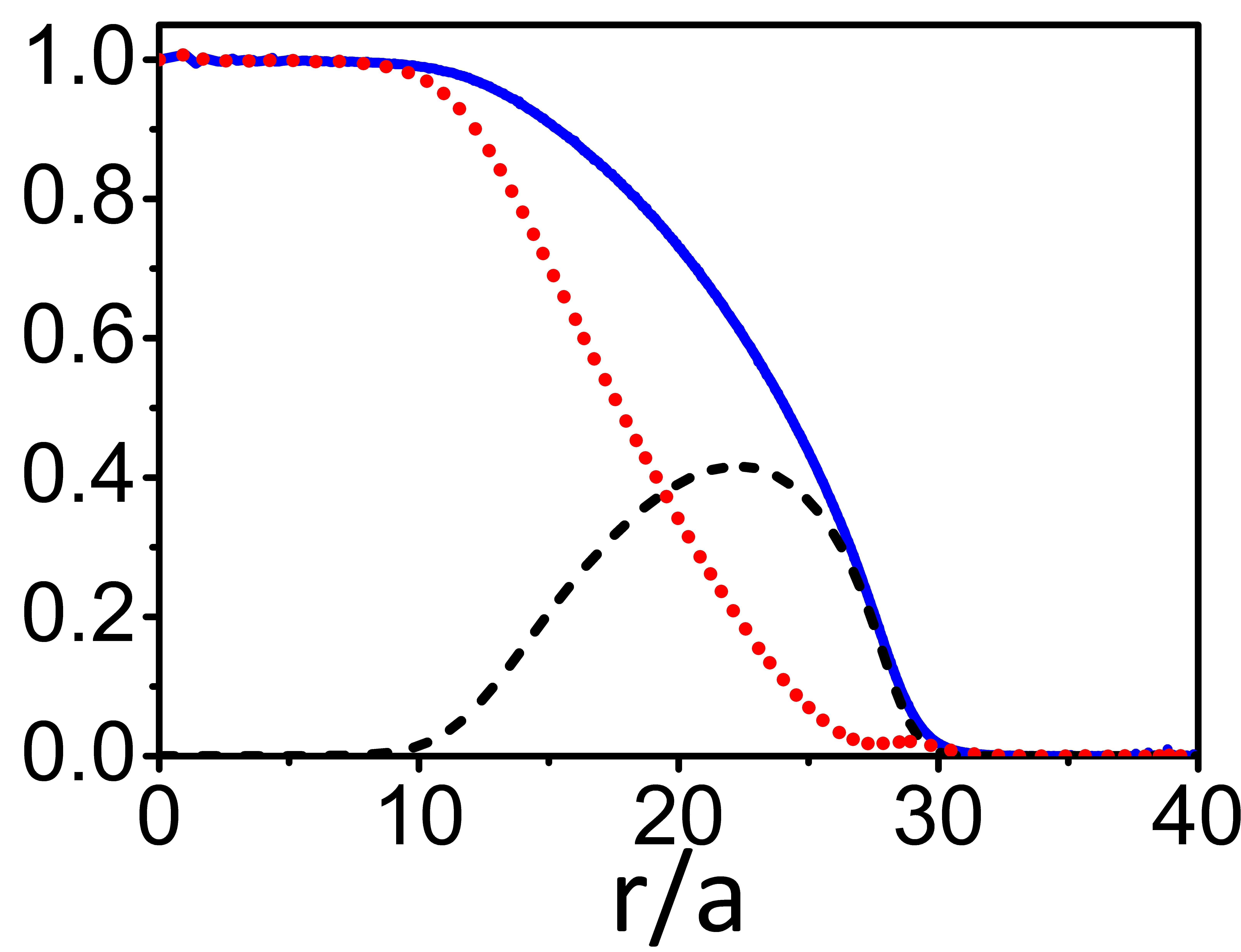} \\
	(a) & (b) & (c) & (d) \\

    \multicolumn{4}{c}{\fbox{Far Field ($\tau \rightarrow \infty$)}} \\	
    \includegraphics[clip,width=39mm, height=25mm, angle=0]{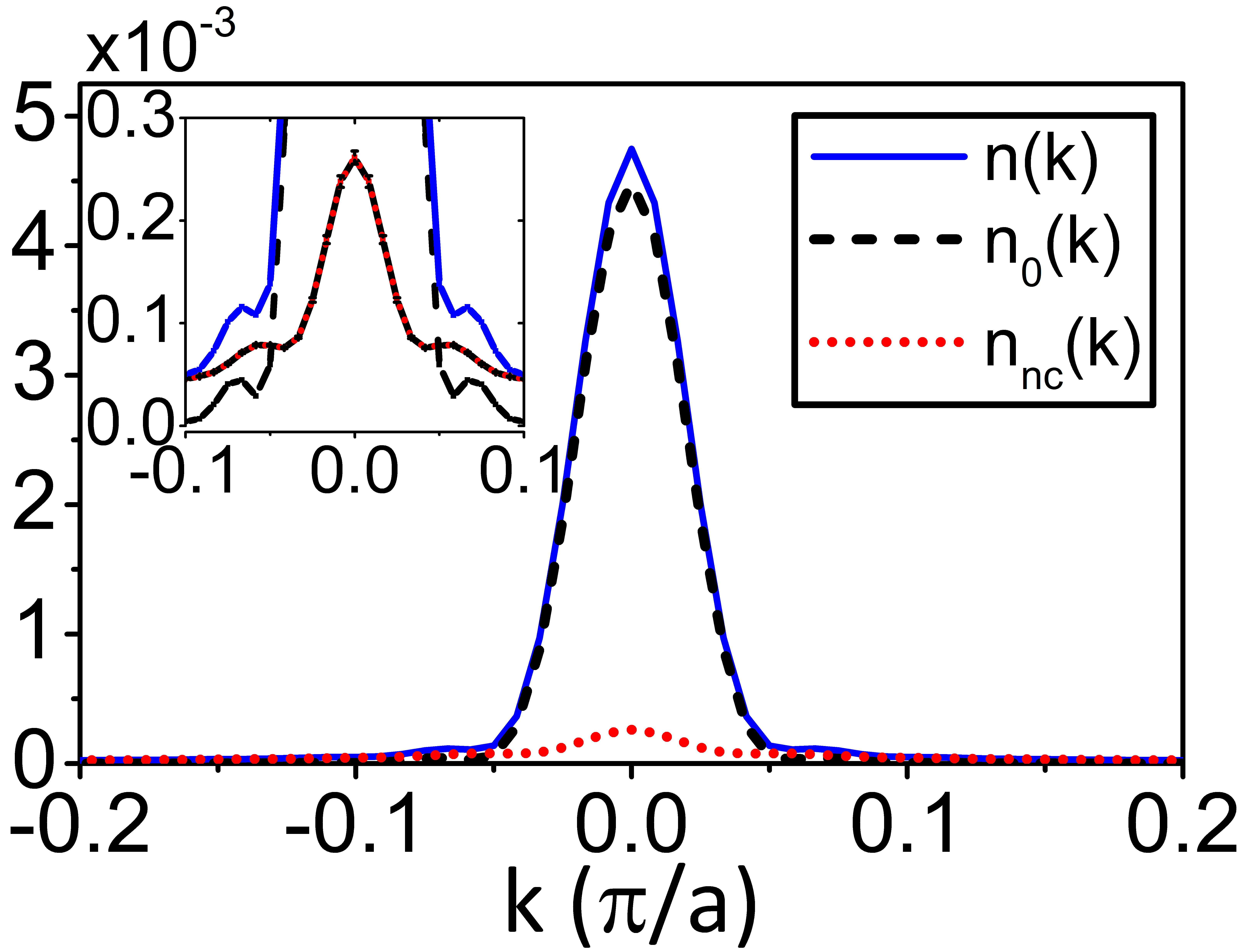} &
    \includegraphics[clip,width=39mm, height=25mm, angle=0]{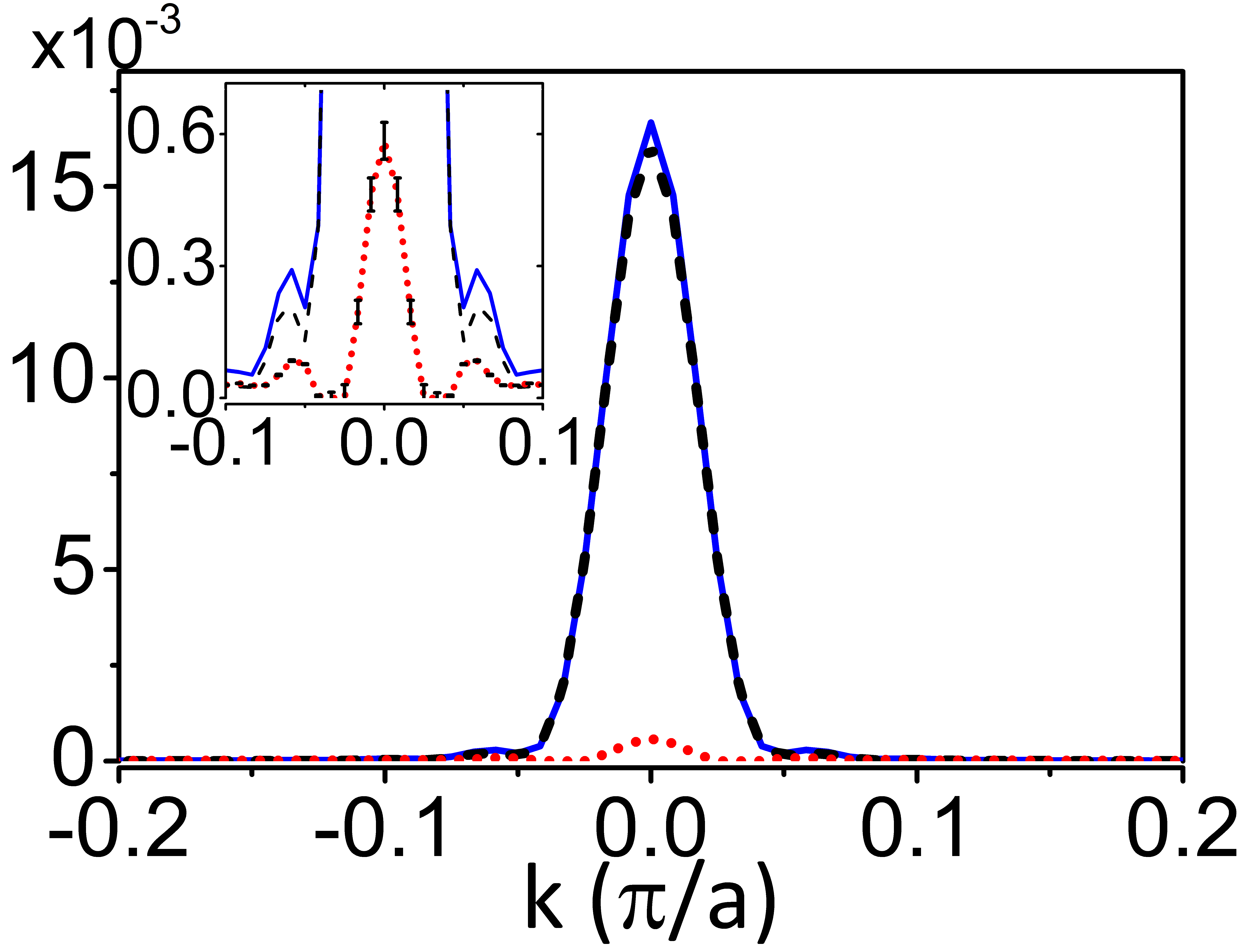} &
    \includegraphics[clip,width=39mm, height=25mm, angle=0]{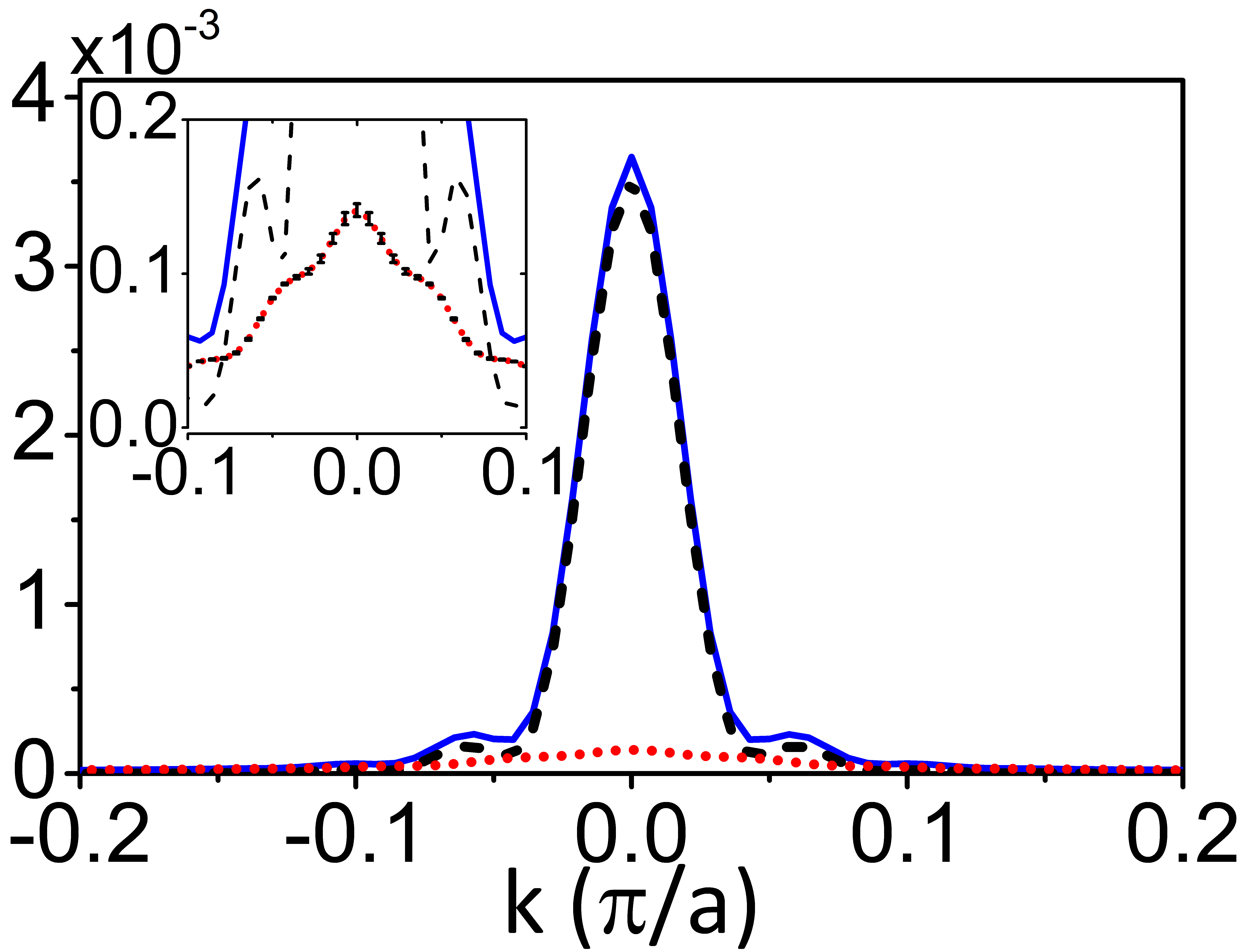} &
    \includegraphics[clip,width=39mm, height=25mm, angle=0]{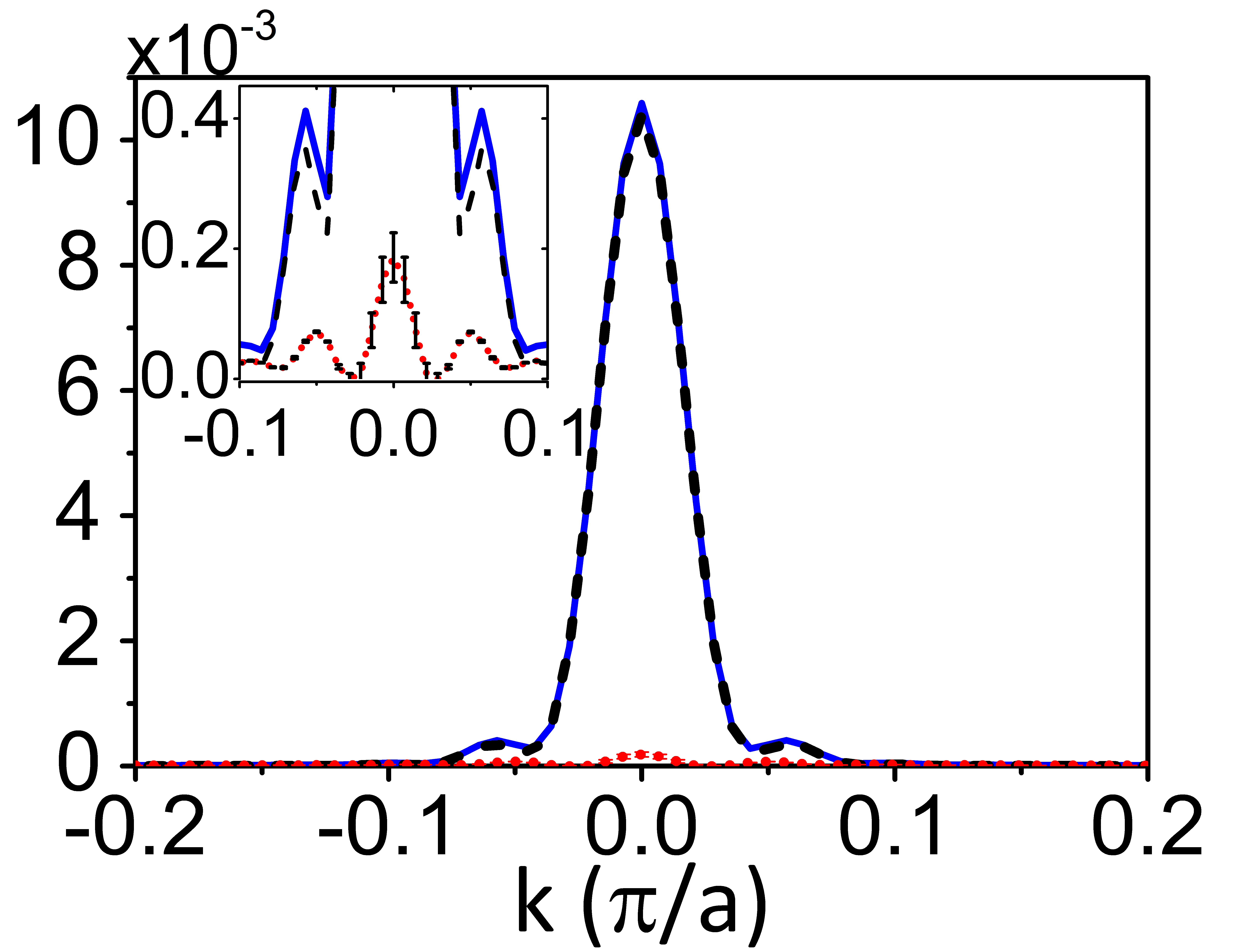} \\
	(e) & (f) & (g) & (h) \\		

    \multicolumn{4}{c}{\fbox{Finite TOF ($\tau = 20$ ms)}} \\
    \includegraphics[clip,width=39mm, height=25mm, angle=0]{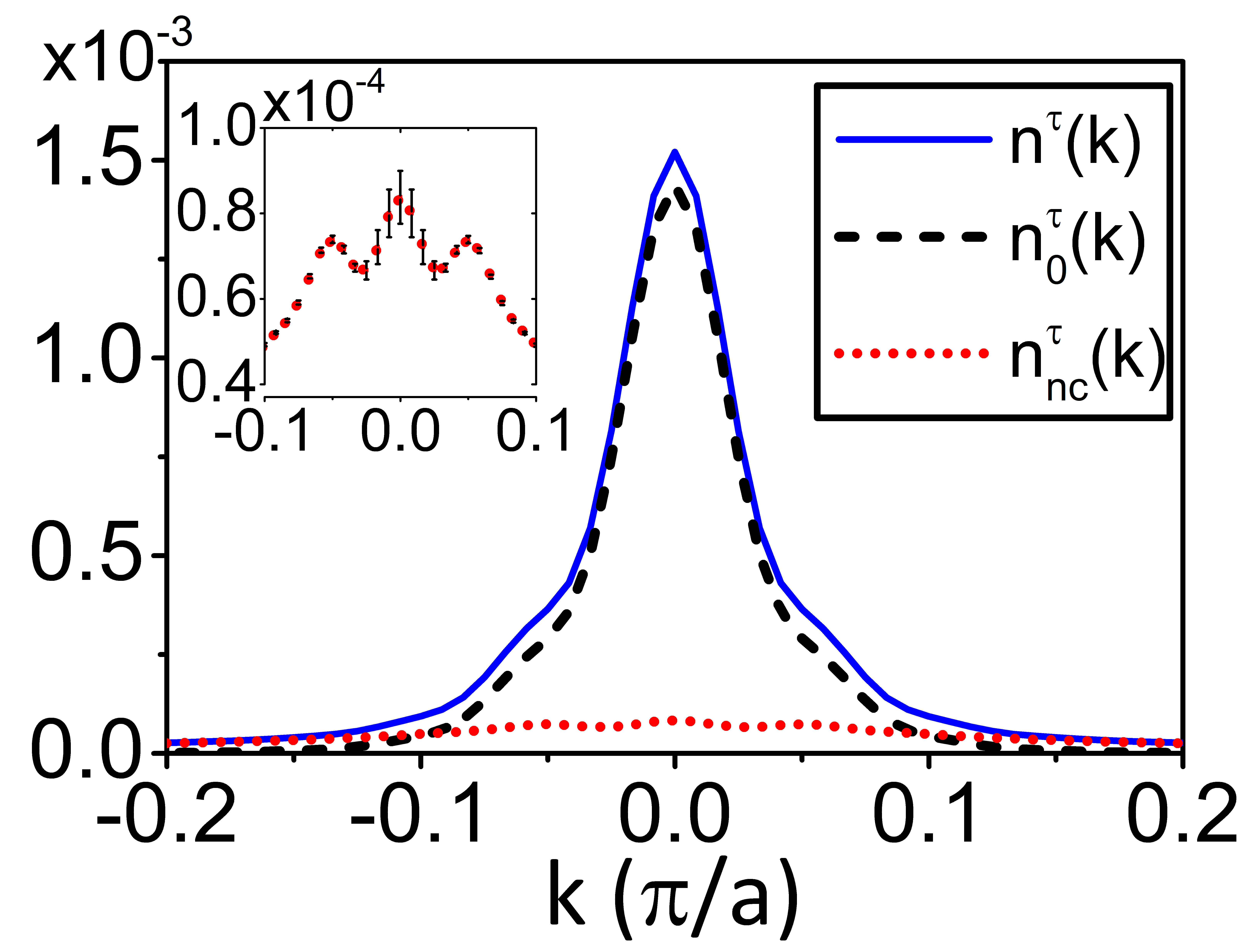} &
    \includegraphics[clip,width=39mm, height=25mm, angle=0]{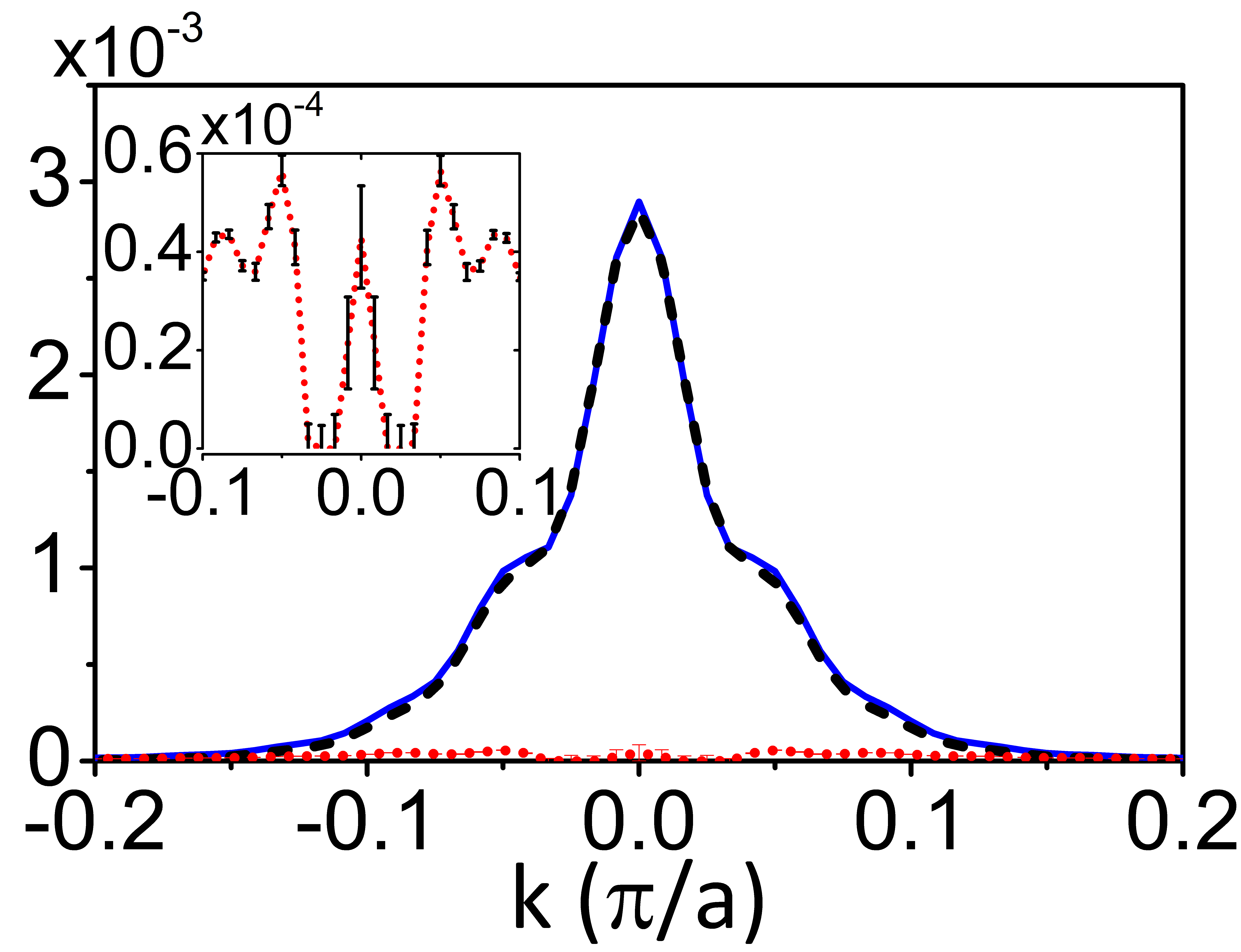} &
    \includegraphics[clip,width=39mm, height=25mm, angle=0]{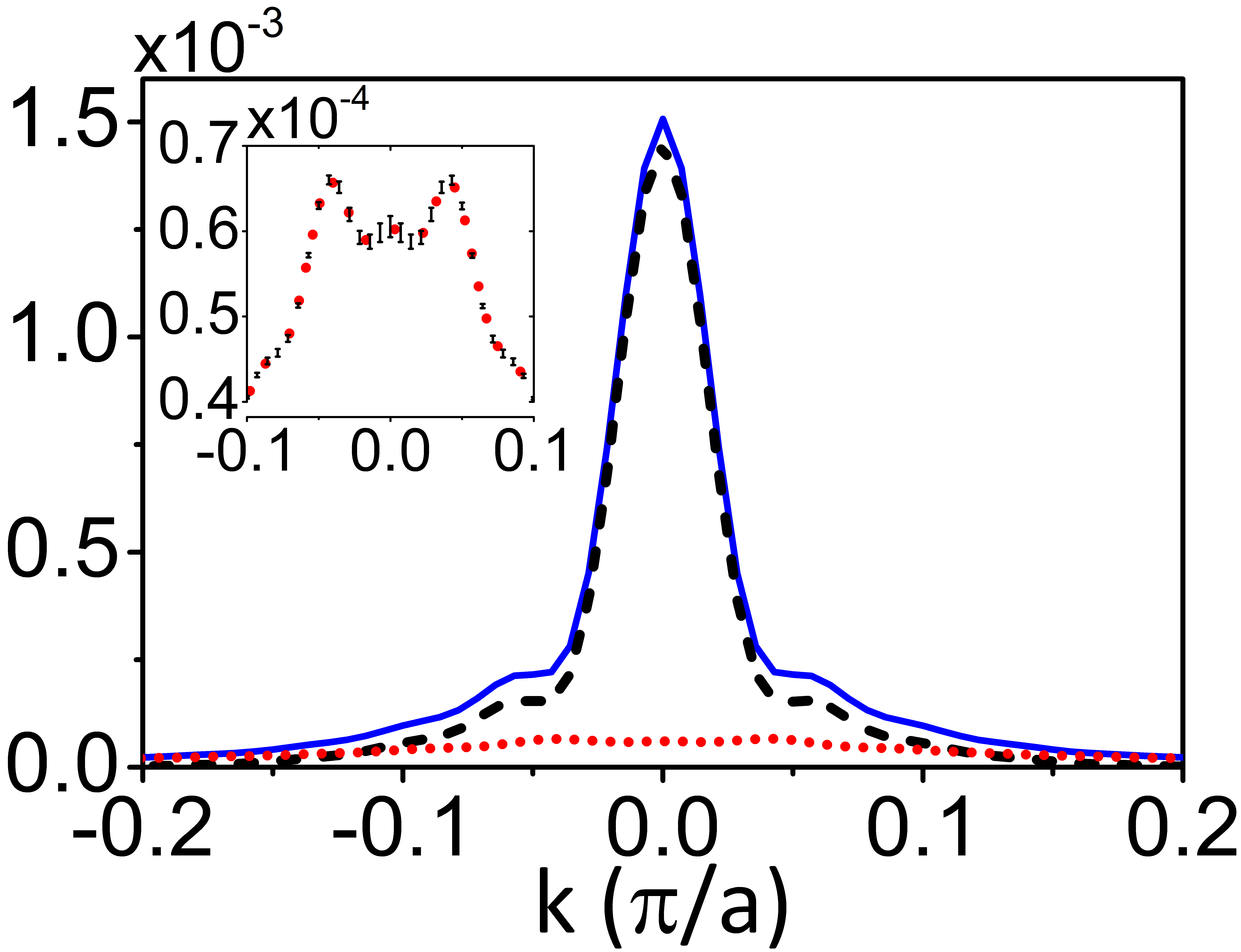} &
    \includegraphics[clip,width=39mm, height=25mm, angle=0]{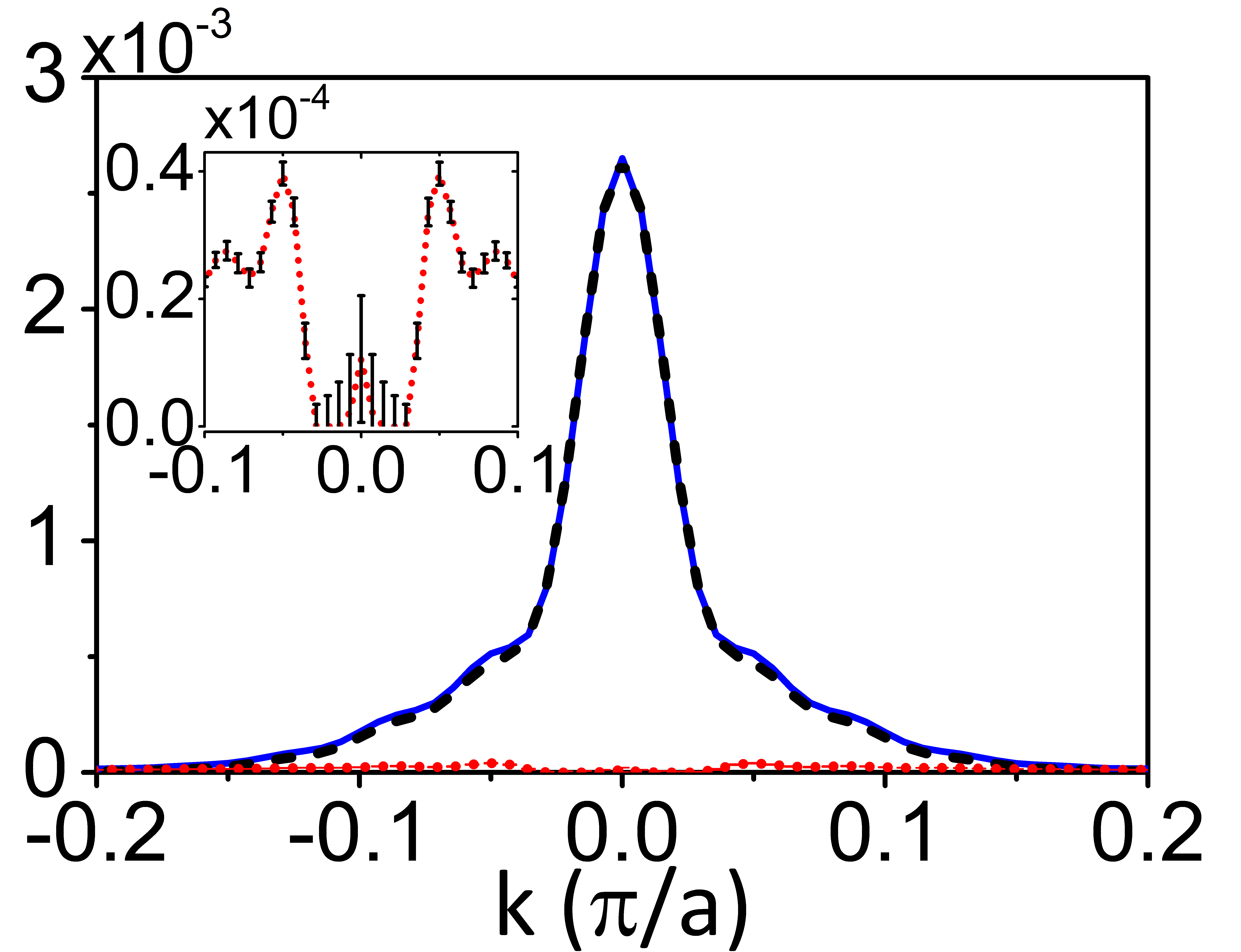} \\
	(i) & (j) & (k) & (l) \\
\end{tabular}
\vspace{-3mm}
\caption{
Top row (a-d) are the spatial particle densities, middle row (e-h) and bottom row (i-l) are the far field column integrated and finite TOF momentum distributions. Insets on the middle line (e-h) show fine features for all distributions and last line (i-l) show zoomed versions of the finite TOF non-condensate distribution ($n^{\tau}_{nc}(k)$). For the momentum distributions in the middle and last line the y-axis is in arbitrary units, but the scale is the same for all images.
} 
\label{Fig:prof}
\vspace{-6mm}
\end{figure*}


In fig. (\ref{Fig:prof}a-d), we present the different density profiles corresponding to large U. (The trap frequency has been adjusted so that $n(r=0)=1$; the density is 1 atom/lattice site at the center of the trap.) We note that the underlying distributions here are exact. At $U/t=40$, the Mott Insulating (MI) domain appears as an integer plateau. Similar distributions obtained in previous studies \cite{rey03,Yi07, anderson04,oosten01} using MFT + LDA cannot capture the interface between the developing MI and superfluid domains that play an increasing role at high $U/t$ since these studies effectively stitch together densities for different radii, while assuming that there are no exchanges across different shells. Therefore, studies using these distributions as masks for fitting purposes must be carefully interpreted. 

The column integrated momentum distributions in the far-field, shown in fig. (\ref{Fig:prof}e-h), exhibit secondary peaks similar to \cite{Svistunov02, wessel04}. Note that these peaks can arise not only due to the finite extent of the condensate but rather all modes (fig. (\ref{Fig:prof}e-h) insets). However, it is only around the MI regime that it will be primarily due to the condensate, and so we see that the peak forms around $k = 2\pi/\xi_0$, where $\xi_0$ is the width of the condensate. This is specifically due to the way the condensate forms between boundaries (in this case between the MI domain and the vacuum). We present statistics of the secondary peak in table ({\ref{Tab:peak2stats}) from which it is evident that its size relative to the central peak and the contribution of the condensate mode to it is much larger (2.8 times at low T and 5 times at high T) when the system has a MI domain. 
 
Finite TOF effects, presented in fig. (\ref{Fig:prof}i-l), alter the far field distributions by suppressing and blurring the central low $k$ values, as expected \cite{Prokofev08}. Higher order modes of \N with rapid spatial variations are not significantly affected by the site dependent phase shift. Thus, the maximum effect is on the condensate distribution. The time scale for the condensate to reach the far field ($\tau_{ff}$) is $\propto R\xi_0(1-\xi_0/2R)$, where $R$ is the radial extent of the condensate. This leads to larger $\tau_{ff}$ for $U/t = 25$ and so for the fixed $\tau = 20$ms, the central peak sees a greater suppression (and surrounding region greater enhancement) than $U/t = 40$. Although, $n(k)$ and $n^{\tau}(k)$ are both broader due to the secondary peaks, the latter case has relatively more condensate atoms. They would {\it not} be captured in fitting schemes used in experiments \cite{trotzky10,mckay_thesis}.  

We note that the broader structure is observable within the experimental resolution. Using $\Delta k = (m\lambda\pi/h\tau) \Delta r/a$, where $\lambda$ the wavelength of the optical lattice, $\tau$ the expansion time, $\Delta r$ the resolution, we obtain the  $\Delta k$ resolution. Using $Rb^{87}$, $\tau = 20$ ms, $\lambda = 800 nm$ and typical resolving power of $\Delta r = 3 \mu m$ gives $\Delta ka \sim 0.026\pi/a$. Features in fig. (\ref{Fig:prof}i-l) are spread over $\Delta k \sim 0.06\pi/a$.

\begin{figure}[b]
  \begin{tabular}{cc}
	{$U/t=3.4$ $\omega=37.7$ Hz} & {$U/t=25$ $\omega = 68.1$ Hz} \\
	\fbox{$k_bT/t=3.93$ $n_0=0.0151(1)$} & \fbox{$k_bT/t=3.27$ $n_0=0.00142(4)$} \\
    \includegraphics[clip,width=39mm, height=25mm, angle=0]{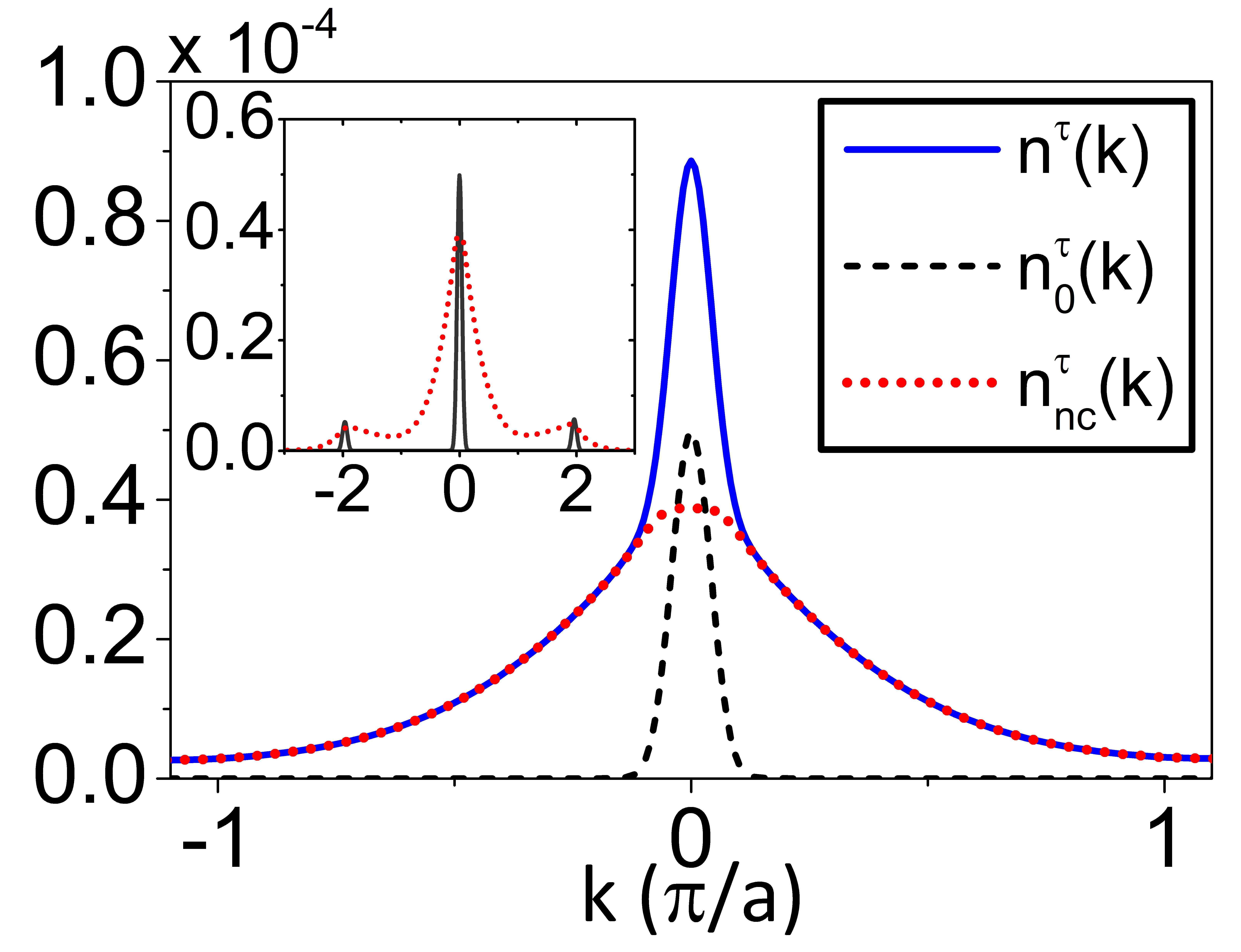} &
    \includegraphics[clip,width=39mm, height=25mm, angle=0]{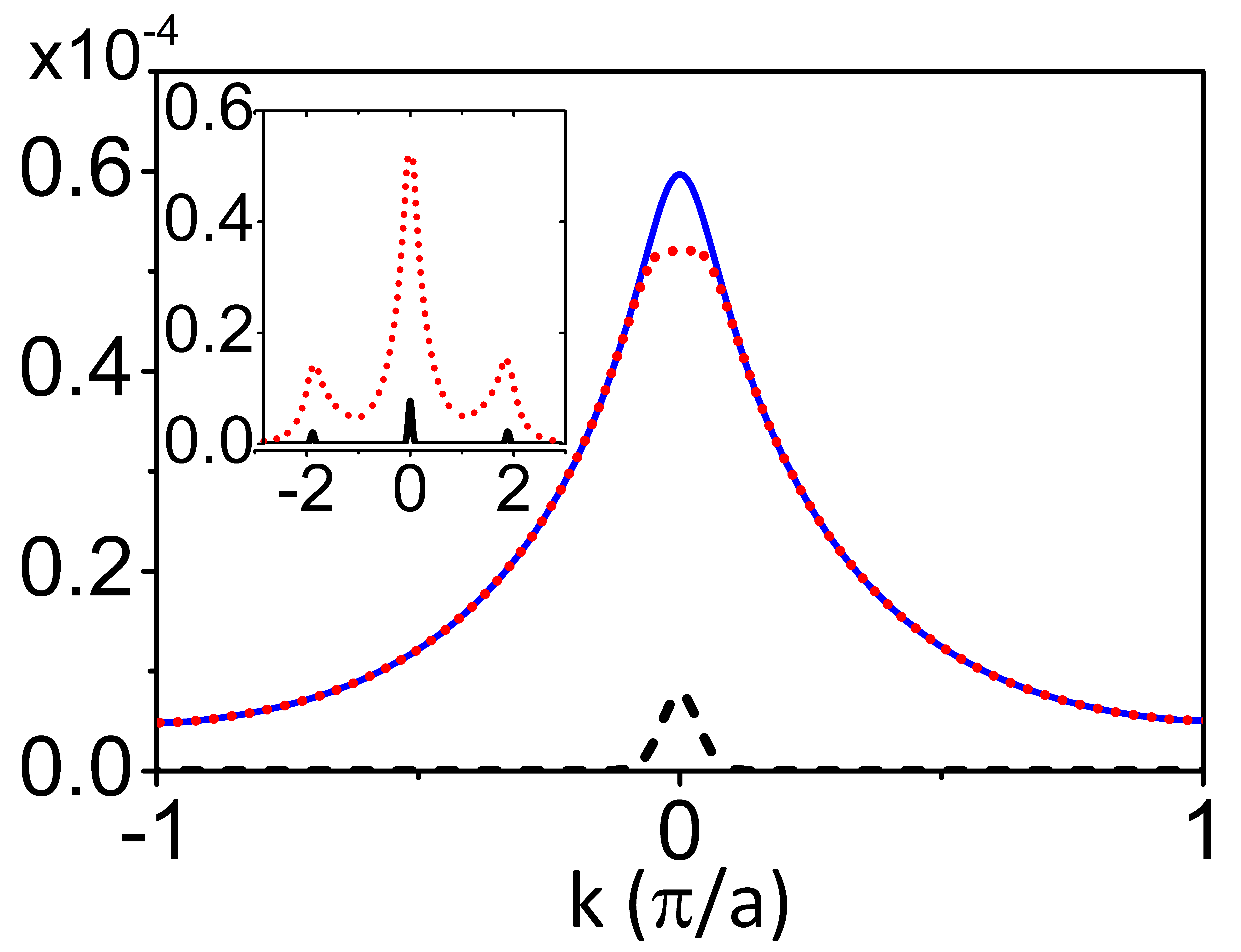} \\
 	(a) & (b) \\ 
 \end{tabular}
 \vspace{-3mm}
  \caption{Closer look at the distributions near the transition temperature for a system of $N \sim 64,000$ atoms. (a) and (b) are finite time TOF distributions with $\tau = 20ms$. Depending on the cut-off $k$ value chosen ($k_l$) there are between $500$ to $5000$ atoms for $U/t = 3.4$ ($k_l = 0.01$,$0.1$) and $300$ to $1260$ atoms for $U/t=25$ ($k_l = 0.035$,$0.075$). The exact number is $960$ and $90$ respectively. The insets show $n_{nc}(k)$ and $n_0(k)$ with 3 peaks.}
\label{Fig:specialprofs}
\vspace{-6mm}
\end{figure}

Fig. (\ref{Fig:specialprofs}a and b) show underlying distributions around the transition temperature ($T_c$) for a weakly interacting ($U/t=3.4$) and a strongly interacting ($U/t=25$) system when $n_{nc} = 0.98(5)$ and $0.998(6)$. In accordance with previous studies, we see strong central peaks \cite{Svistunov02,trivedi08} even at these warm temperatures. Furthermore, the central and surrounding Bragg-peaks are not representative of the condensate on its own. The magnitude of the error due to approximate fits is specific to the fitting procedure used. However, we can systematically analyze boundaries of possible errors due to the issues we discussed above. We estimate the coherence fraction by calculating the ratio of atoms under the central peak (up to a limit $k_l$) to the total number of atoms. (The numbers are calculated from column integrated images without the wannier envelope so that we do not worry about its subtraction.) The range is chosen large enough to accommodate finite optical resolvability in experiments. It should also allow for related fitting procedures. 

\begin{figure}[t]
  \begin{tabular}{cc}
	\fbox{$U/t=25$ $\omega=68.1$ Hz} \\
	 \vspace{1mm}
    \includegraphics[clip, height=40mm, width=73mm, angle=0]{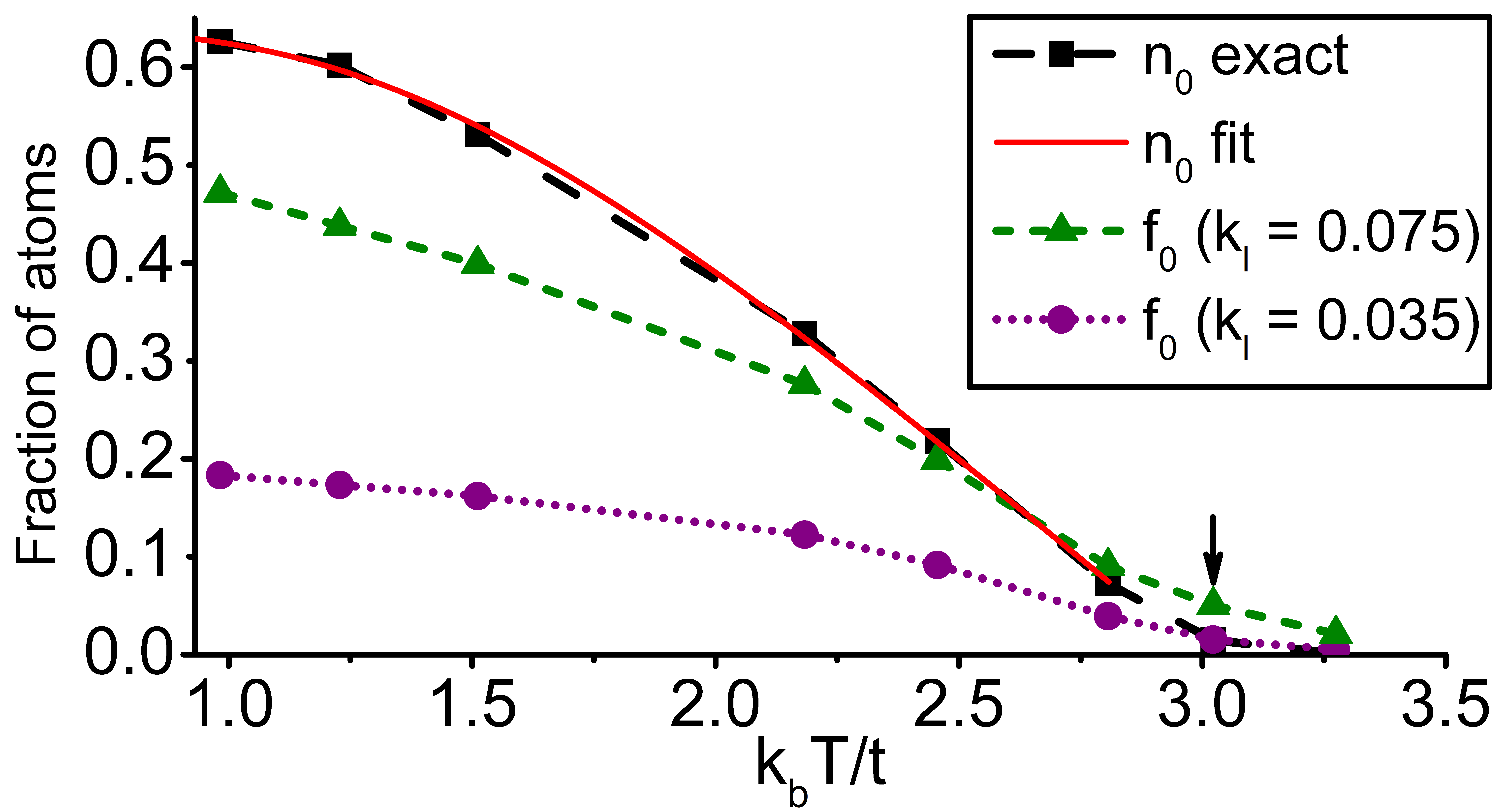} \\
	(a) \\
	 \fbox{$U/t=40$ $\omega = 67.6$ Hz} \\	
    \includegraphics[clip, height=40mm, width=73mm, angle=0]{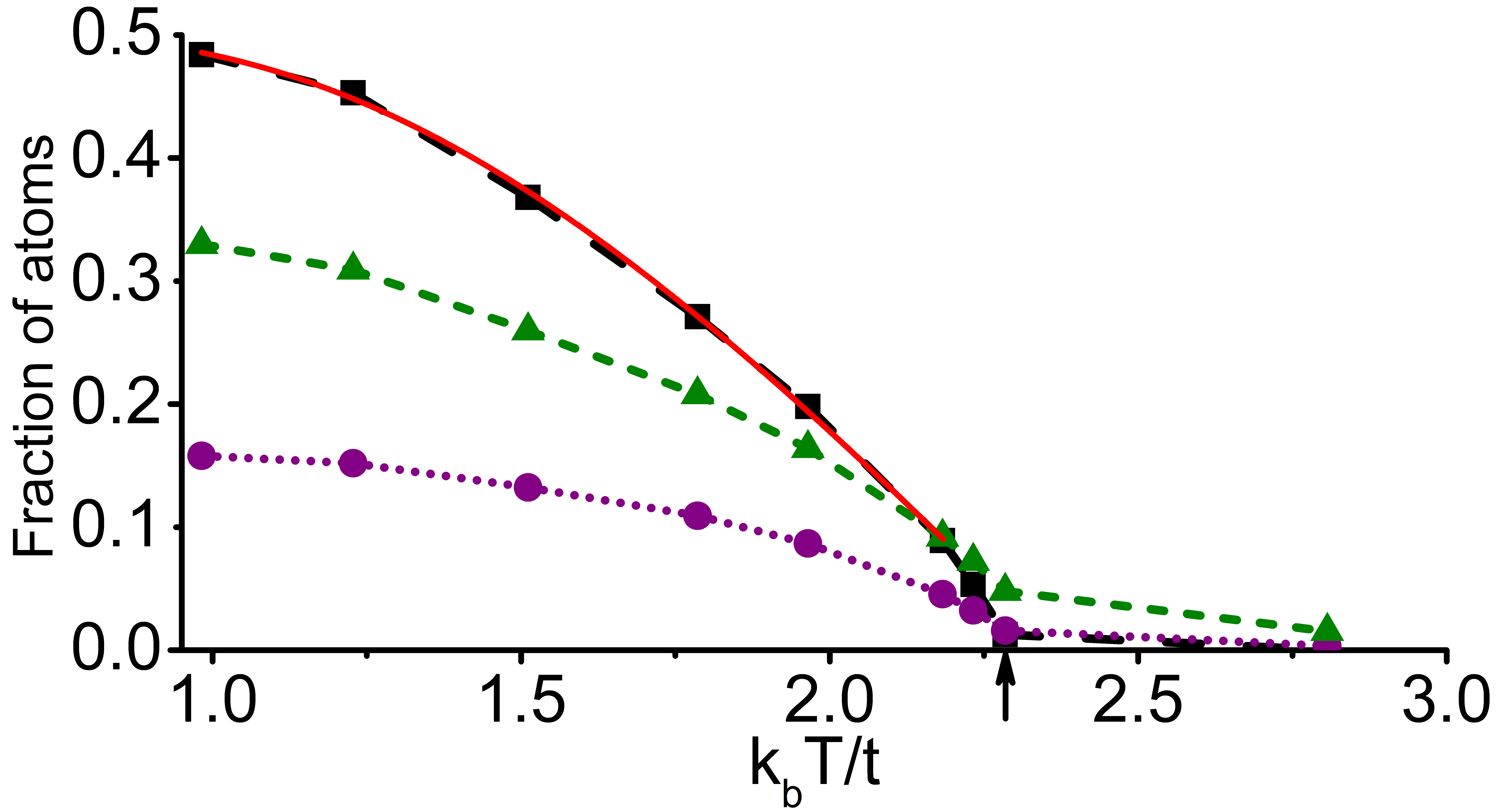} \\
 	(b)  \\   
 \end{tabular}
 \vspace{-3mm}
  \caption{Condensate fraction ($n_0$) as a function of temperature (T) compared with coherence fraction ($f_0(T)$) measurements for different cutoff $k_l$. ($f_0 \equiv \sum_{|k|<k_{l}}n_0(k)/N$.) We have fit the $n_0$ to the functions given by (\ref{eqfit}) where $n^{*}_0 = 0.638(7)$ ($0.502(6)$), $k_bT_i/t = 2.99(2)$ ($2.36(1)$) and  $g = 1.91(9)$ ($2.4(1)$) for $U/t = 25$ ($40$). The short arrows indicate $k_bT_c/t = 3.023$ $(2.285)$ for $U/t = 25$ ($40$) where $n_0 \sim 0.01$}
\label{Fig:n0vsT}
 \vspace{-3mm}
\end{figure}

We present comparisons with exact results in fig. (\ref{Fig:n0vsT}a and b) approximated by:
\vspace{-2mm}
\begin{equation}
n_0(T)=n^{*}_0(1 - exp(g(1-\frac{T_i}{T})),
\label{eqfit}
\end{equation}
away from the critical regime \cite{footnote2}. We see that at low T, the depletion is severely overestimated whereas around the transition it is underestimated. It is worth noting that general trend of the error is in accordance with $f_0$ measurements presented in \cite{mckay_thesis}, where it is further exacerbated by experimental noise and  additional fitting errors due to inaccurate background subtraction and possible interaction effects during expansion of the gas. 

Care is needed in the estimation of $T_c$ using $f_0$. In the general case, extrapolating from our data, for $k_l\geq0.035$, $f_0(k_l) > n_0$. If $T_c$ is to be estimated correctly, $|\partial^2 f_0/\partial T^2| > |\partial^2 n_0/\partial T^2|$ is required for rapid convergence to the same $n_0(T)\rightarrow0$. Futher, following \cite{Vicari09}, a study of the trap-size scaling behavior analogous to finite size scaling studies done for homoegeneous systems would be needed to exactly identify critical exponents and $T_c$ for trapped systems. Here we use a simpler working definition $n_0(T_c) \sim 0.01$ that suffices to show that estimates of $T_c$ using $f_0$ would be erroneous: in fig. (\ref{Fig:n0vsT}a and b), $k_bT_c/t = 3.022$ $(2.284)$, but using $f_0$, $3.023 (2.29)<k_bT_c/t<3.3 (2.81)$ for $U/t = 25$ $(40)$. Furthermore, it should also be clear that a smooth change in $f_0(T)$ across the transition might make it impossible to use the inflection point as an accurate estimate of $T_c$. 

A simple way for experiments to obtain the correct fitting function might be to calibrate the coldest possible measurements of $f_0$ against $n^{*}_0 \sim n_0(T=0)$, the saturated $n_0$ from QMC. The approximate temperature could be approximated from entropy measurements as in \cite{mckay_thesis,pasienski10}. The TOF image used to estimate $f_0$ should be then fitted correctly: the bi-modal structure of the condensate seems amenable to be fit by a narrow Thomas-Fermi like function plus a broader Gaussian. Fitting the non-condensate correctly is somewhat easier since, at low T the height of $n_{nc}(k)$ at low k is small and fluctuations can be integrated out. On the other hand, at high T it is flat under the peak and cannot be assumed to be gaussian-like outside. 


We have presented the effects of interactions on the components of the spatial and momentum distributions of bosonic particles in a trapped optical lattice. Our unbiased estimates of \N elucidate the potential problems in mapping between the coherence fraction and the exact \CFe, and how to account for them. Using exact distributions for QMC, experiments will get more accurate estimates of \CFe. In a future work, we will study the effects of interaction during TOF and present comparisons with experimental systems at higher density \cite{ray12b}. We will also explore the entropy (S) mapping using QMC that is need to compare with $n_0(S)$ from experiments. The method we have presented in this paper will be crucial to such studies.  

We wish to thank Brian DeMarco, David McKay, Fei Lin, Vito Scarola, Norm Tubman and Matthias Troyer for useful discussions. This work is supported by the DARPA-OLE program. Computation time was provided by the XSEDE resources at NCSA (University of Illinois in Urbana-Champaign) and TACC (Texas). 

\let\clearpage\relax

\section{Tables}
\vspace{-6mm}
\begin{table}[h]
\begin{tabular}{|c|c|c|c|c|c|c|}
\hline
$U/t$ & $k_bT/t$ & $n_0(k_{p2})/n(k_{p2})$ & $n_0(k_{p2})/n(0)$ \\
\hline
\multirow{2}{*}{$25$} & $2.46$ & 0.405 & 0.0098\\
& $0.98$ & 0.742 & 0.0127\\
\hline
\multirow{2}{*}{$40$} & $1.96$ & 0.696 & 0.0414\\
& $0.98$ & 0.86 & 0.034\\
\hline
\end{tabular}
\caption{
Statistics of the secondary peaks: fraction of condensate at the maxima of the secondary peak ($k_{p2}$) and size relative to the central peak.  
}
\label{Tab:peak2stats}
\vspace{-6mm}
\end{table}


\begin{thebibliography}{90}
\bibitem{footnote2} We find that the function (\ref{eqfit}) is a good fit for our QMC data provided $k_bT/t \leq 2.183$ $(2.807)$ where $n_0 \geq 0.0891(4)$ $(0.0726(3))$ for $U/t = 25$ $(40)$.
\end{thebibliography}
\end{document}